# Standardised Reputation Measurement

**Draft - please do not cite without prior permission from the author**

**Peter Mitic** [1,2,3]

[1] Santander UK
 2 Triton Square, Regent's Place, London NW1 3AN
 Email: peter.mitic@santandergcb.com
 Tel: +44 (0)207 756 5256

[2] Department of Computer Science, University College London
 Gower Street, London WC1E 6BT

[3] Laboratoire d'Excellence sur la Régulation Financière (LabEx ReFi), Paris

## Abstract

*Well-defined formal definitions for sentiment and opinion are extended to incorporate the necessary elements to provide a formal quantitative definition of reputation. This definition takes the form of a time-based index, in which each element is a function of a collection of opinions mined during a given time period. The resulting formal definition is validated against informal notions of reputation. Practical aspects of data procurement to support such a reputation index are discussed. The assumption that all mined opinions comprise a complete set is questioned. A case is made that unexpressed positive sentiment exists, and can be quantified.*

## Disclaimer

# 1. INTRODUCTION

Reputation measurement has gained increasing prominence in recent years as organisations become much more aware that their reputation matters. Previously a somewhat subjective concept, a formal definition of reputation, as well as its measurement, have been elusive until very recently. The aim of this paper is to provide a clear distinction between reputation, and two closely related concepts - sentiment and opinion – and then to formalise the definition of reputation in quantitative terms.

## 1.1 Informal Definitions for Sentiment and Reputation

The Oxford English Dictionary defines sentiment as ``A view or opinion that is held or expressed`` (https://en.oxforddictionaries.com/definition/sentiment). Other dictionaries concur. Liu (2015) uses the word *sentiment* in the sense of a positive or negative *feeling*, and also introduces the word *opinion* to indicate a broad context covering sentiment (Liu definition), evaluation, appraisal, attitude and associated information such as the opinion holder and the opinion target.

Reputation is a collection of opinions, but there is more to it than that. Reputation expresses the relationship between opinion holders and the performance of the target, referred to the expectation of the opinion holder. An informal definition is given in Mitic (2017), but is amended below to be consistent with the Liu definition of opinion. Therefore we formulate these informal definitions.

- *Senti*m*ent*: A view that is held or expressed
- *Opinion*: Sentiment expressed by a holder of a target at a particular time
- *Reputation*: Collective opinions, established over time, that can conflict with the expectation that the opinion holders have of the target

This informal definition of reputation encapsulates the idea that reputation is a difference between "what you expect and what you get". It corresponds broadly to the definition of reputation from the Basel Committee (BCBS157 2009).

## 1.2 Formal Definitions for Sentiment, Opinion and Reputation

The following formal definitions of sentiment and opinion are based on the definitions in Liu (2015). Liu defines three separate components of sentiment: polarity (positive, negative or zero), intensity, (a measure of the extent of polarity – numeric or otherwise) and type (rational or emotional). In practice, polarity and intensity emerge from data mining as a single entity, and it more convenient to subsume the type into a more general categorisation vector *C* (see below). Consequently, we will define a <<standard measure of sentiment>>, *S*, as a real number in the range [-1, 1].

$$S \in \mathbb{R}: \; -1 \leq S \leq 1 \qquad (1)$$

An account of methods which may be used to quantify sentiment may be found in, for example Jurafsky (2008) or Bishop (2007). They include methodologies such as Naive Bayes, Artificial Neural Nets and Support Vector Machines.

Opinion, $\boldsymbol{O}$, is minimally a 5-dimensional vector which extends the concept of sentiment to include, a unique identifier $i$, a timestamp $t$, a target $G$, the sentiment $S_i$ (indexed to its identifier since it is often used in isolation) and its holder (originator) $H$. It is useful, although not necessary, to add a sixth component to $\boldsymbol{O}$: a categorisation vector $\boldsymbol{C}$, which can be used to classify the opinion (for example, as social/business, influence level of the holder …). This component is useful for analysis of factors that affect reputation, and subsumes Liu's sentiment "type".

$$\boldsymbol{O} = \{i, t, G, H, S, \boldsymbol{C}\} \qquad (2)$$

Reputation at time $t$ can then be defined in terms of a collection of opinions $\{\boldsymbol{O}\}_{t \in T}^{i \in I}$ \, where $T$ is an indexing set for time and $I$ is an indexing set for unique identitifiers. The reputation of an organisation $G$ at time $t$, $R_G(t)$, can then be given as some generic function $\rho$ of the collection of opinions:

$$R_G(t) = \rho(\{\boldsymbol{O}\}_{t \in T}^{i \in I}) \qquad (3)$$

The function $\rho$ could be, for example, a weighted average of the sentiments expressed in the collection of opinions. With this approach, if $w_i$ is the weight assigned to the opinion $\boldsymbol{O}$ with unique identifier $i$ and sentiment $s_i$, and the indexing set $I$ has $n$ elements, the function $\rho$ and the reputation at time $t$ take the form

$$R_G(t) = \frac{\sum_{i=1}^{n} s_i w_i}{\sum_{i=1}^{n} w_i} \qquad (4)$$

The informal definition of reputation in the previous sub-section includes the reference "established over time". To extend the definition $R_G(t)$ to cover the times in the indexing set $T$, we define the reputation, $\hat{R}_G$, of the target $G$ as the time series

$$\hat{R}_G = \{R_G(t)\}_{t \in T} \qquad (5)$$

This definition depends on reputation measurements taken over an extended period. It is not sufficient to deal with cases where a potential opinion holder notes a small number (perhaps one only) of isolated comments, and formulates his/her own opinion based solely on that.

## 2. DATA MINING FOR REPUTATION

In this section we give a brief overview of the practicalities of the data mining processes needed to implement viable reputational analysis. Full details may be found in Mitic (2017). The process is, for a given time period (typically 24 hours):

1. Receive `contents' (corresponding to equation (2)) by electronic feeds from relevant public sources of opinion: news reports, radio and TV broadcasts, press releases, reports from trade events, comments on social media (Twitter, Facebook, blogs etc.).
2. Analyse each content for sentiment, define a weight (e.g. to reflect the influence of the opinion holder), resulting in a standardised sentiment (equation (1)) for each.

3. Compose a reputation index component using all content received in the time period (equation (4)}, or more generally (3)).

A reputation index, as defined in equation (5) can then be compiled by accumulating the results of the above process for a sequence of intervals *t* in a set *T*. Figure 1 summarises the reputation index procurement process.

Figure 1: Reputation Index Procurement Process

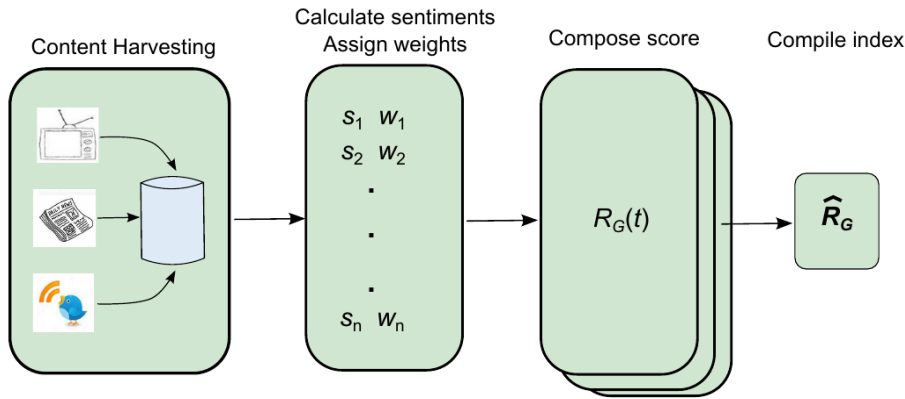

The sequence shown in Figure 1 is intended to remove subjectivity from the reputation index procurement process. This is possible provided that the sources for the data mining stage form a complete and comprehensive set and that the analysis of the contents is sufficient to determine sentiment accurately. The completeness assumption will be questioned in the next section. The definitions of reputation given can be said to induce bias by not quantifying contents that have not been received.

The possible existence of `missing' positive sentiment is summed up in a quote by Donald Rumsfeld, who was US Secretary of Defence in the George W. Bush administration (NATO 2002): "*The message is that there are no "knowns." There are things we know that we know. There are known unknowns. That is to say there are things that we now know we don't know. But there are also unknown unknowns. There are things we don't know we don't know.*". "Missing" positive sentiment corresponds to "unknown unknowns".

## 3. NEGATIVE OPINION BIAS: METHODOLOGY

The definitions in equations (3) and (4) apply to all contents that are received. It clearly cannot be applied to comments that are not received. The comment that any reputation measurement cannot be accurate has been heard informally. The reason is bias towards negative sentiment, summed up in the phrase ``No news is good news". Agents (individuals and groups/corporates may be forthright in expressing strong sentiment – either positive or negative. They might also express, relative to the ambient sentiment (i.e. the mean reputation score), very mild negative sentiment if they are slightly annoyed, but might not bother with corresponding positive sentiment if they are only mildly satisfied. Therefore some contents that express positive sentiments might be 'missing'.

There are indications that such sentiment bias exists. Aktolga and Allan (2013), and Cook and Ahmad (2015) both detected positive and negative sentiment bias. Kelly and Ahmad

(2015), found statistically significant predictive power due to only negative sentiment when predicting stock market returns.

### 3.1 Bias Measurement

To try to estimate how many positive sentiments might be missing, we look specifically at reputation scores near the truncated mean (i.e the mean excluding the largest and the smallest - avoid extremes), $m$, of a set of reputation scores. This reflects the idea that sentiments that are mildly above average might be `missing'. The argument proceeds first by considering the cumulative reputation score $S_T$ over a given time period (i.e. the sum of the elements in $\hat{R}_G$ in equation (5)).

$$S_T = \sum_{t \in T} R_G(t) \qquad (6)$$

Figure (2) shows the three typical cumulative reputation score profiles: positive, negative and zero trending. In each case, the cumulative reputation score from time 0 to $t$, $S_{[0,t]}$, is plotted against time. With positively trending cumulative reputation, 'missing' positive sentiment is measured by seeking a negative trend in a neighbourhood of $m$. For a negative trending cumulative sum, we reverse the sign of the scores, and continue as though it were positive trending.

Figure 2: Typical examples of Cumulative Reputation Scores

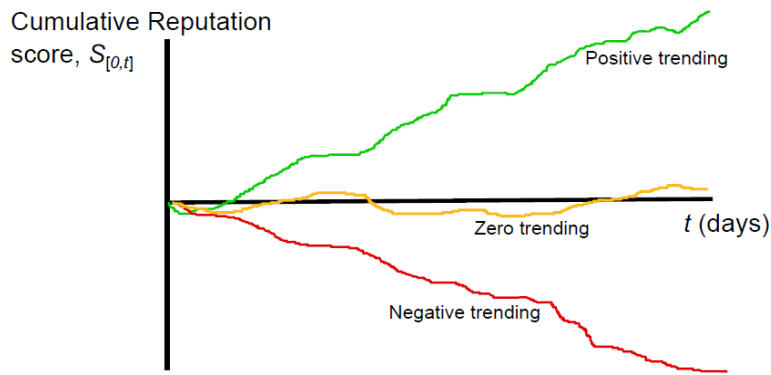

First we compare the skewness of scores in a neighbourhood of $m$ with the skewness of all the scores. To do this we calculate the skewness of elements $R_G(t)$ in the range $[m-w, m+w]$ where $w$ is the semi-width of a band, calculated as a percentage of the range of scores (i.e. the difference between the maximum score and the minimum score). A counter skew in a neighbourhood of $m$ is a pointer to `missing' positive sentiment. Skewness is not a useful measure for quantifying the extent of `missing' positive sentiment as the numeric skewness values are not directly related to numbers of contents. Counting reputation scores is.

To estimate the number of missing contents that express positive sentiment, we calculate the *number* of elements in the range $[m-w, m+w]$. A counter trend is measured by calculating the two ratios $\alpha$ and $\beta$ in equation (7), in which $N_w^+$ and $N_w^-$ are the number of scores in the intervals $(m, m+w]$ and $[m-w, m)$ respectively, and $N^+$ and $N^-$ are the number of positive and negative scores respectively.

$$\alpha = \frac{N_w^-}{N_w^+} \; ; \; \beta = \frac{N^-}{N^+} \tag{7}$$

Figure (3) shows the numbers $N_w^+$, $N_w^-$, $N^+$ and $N^-$ for normally distributed scores $R_G(t)$. Such Normal distributions are quite typical for reputation scores.

Figure 3: Normally Distributed Reputation Scores: positively trending cumulative sum

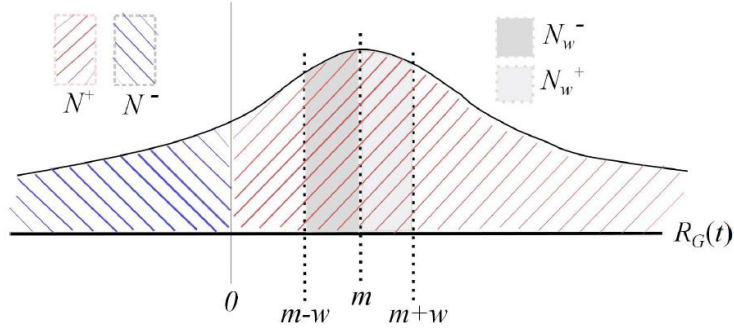

The case α > 1 represents the counter trending case: there are more negative sentiment than positive in a neighbourhood of the mean score, despite an overall positive trend. We compare this ratio with β. We define a measure of the percentage of missing positive sentiments, $M$, by

$$M = 100 \left( \frac{\alpha - \beta}{\beta} \right) \tag{8}$$

## 4. NEGATIVE OPINION BIAS: RESULTS

The data used in the calculations of this paper comprise values of the reputation index for ten UK retail banks, originating from the business intelligence consultancy Alva (www.alva-group.com). For confidentiality reasons, they are labelled Bank1, Bank2, ... Bank10. Alva's reputation index from January 2014 to December 2015 has been used, scaled linearly from its native range [1,10] to [-1,1]. In addition, the scores for the ten banks has been averaged on a per-day basis to produce a reputation score representing retail banking.

The values of $w$ used reflect the target range of sentiment that correspond to the idea that contents that express "small positive" sentiment, relative to the ambient sentiment, are "missing". These cover one third of the range of score values, apart from the 5% nearest to zero, the results from which are unstable. So we take $w$ in a window $W = [2.5\%, 16.5\%]$.

### 4.1 Skewness Results

When there is a positive trend in cumulative reputation, the skewness corresponding to $w$ in the window $W = [2.5\%, 16.5\%]$ is contrary to the overall skewness of the data. When there is a negative trend a contrary skewness is absent. The inference is that there are missing positive sentiments for positive trending cumulative reputation, but not if there is a negative trend. Table 1 shows the results.

Table 1: Negative Sentiment Bias Skewness Results: skew-W is the skewness in the window *W* and Skew-All is the skewness of all the data

| Bank | Skew-W | Skew-All | Trend |
|---|---|---|---|
| 1 | -0.34 | 0.2 | positive |
| 2 | 1.7 | 0.48 | negative |
| 3 | -1.44 | 0.55 | positive |
| 4 | 0 | 0.13 | positive |
| 5 | 1.87 | 0.61 | negative |
| 6 | -1.24 | 0.11 | positive |
| 7 | 1.96 | 0.18 | negative |
| 8 | -0.48 | -0.01 | positive |
| 9 | -0.07 | 0.22 | positive |
| 10 | 2.16 | 0.28 | negative |
| Mean | -0.53 | 0.04 | positive |

### 4.2 Negative Sentiment Bias Measurement Results

A count of reputation scores is more useful for quantifying the extent of `missing' positive sentiment than skewness because it is expressed in terms of actual numbers. Table 2 shows the calculated values of $M$ (equation 8) for each bank that shows a positive trending cumulative reputation score as $w$ varies in the range indicated. Those are the ones that, according to the skewness analysis, have "missing" positive sentiment.

Table 2: Negative Sentiment Bias Measurement Results

| Bank | M |
|------|-------|
| 1 | 11.85 |
| 3 | 23.59 |
| 4 | 4.99 |
| 6 | 7.45 |
| 8 | 3.48 |
| 9 | 13.57 |
| All | 4.28 |

The results in table Table 2 indicate a high dependence on the data used. Consequently, the "All" result "4-5%" should be used to represent "missing" positive sentiment. The highest value results are for the banks that exhibit the steepest positive cumulative score trends (i.e. they have the highest reputation). The opposite effect has been noted in the context of the effect of all banks on the reputation of any particular bank: the ones with extreme reputations (either positive or negative) are influenced the least (see Mitic 2017a for details).

## 5. DISCUSSION

The formal definition of reputation presented in this paper reflects an informal view that reputation refers to collective opinion on the difference between what happens and what is expected. In order to test the view that agents do not always express positive sentiment, we have considered the skewness of the reputation score distribution in a neighbourhood of the mean score. There is some evidence for the existence of "missing" contents that express positive sentiment, as shown by skewness in the neighbourhood of the mean score, counter-trending the overall trend. However, such evidence depends on the measure used, so one should remain sceptical. If the view is that such contents are, indeed, missing, we conclude that:

- "Missing" positive sentiment exists for positive trending cumulative sentiment, but not for negative.
- The reputation score should be inflated by 4-5% to account for such `missing' sentiment.

Having quantified "missing" contents, we mention that contents that are there but should not be, are not rejected. These include malicious contents and contents deliberately generated to affect the reputation score. The problem of quantifying "missing" sentiment could be tackled by survey, although sampling bias would have to be considered carefully.

## 5.1 Acknowledgement

I am grateful for the support of *alva*-Group for granting permission to use their data, discussing their methodology, and for their interest and assistance in the preparation of this paper.